\documentclass[preprint]{aastex}

\begin{document}

\title{Mapping the Galactic Halo. V. Sgr dSph Tidal Debris $60\degr$ 
from the Main Body}

\author{Robbie C. Dohm-Palmer\altaffilmark{1}}
\affil{Astronomy Department, University of Michigan, Ann Arbor, MI 48109}
\email{rdpalmer@astro.lsa.umich.edu}

\author{Amina Helmi}
\affil{Max Planck Institut f\"ur Astrophysik, Garching bei M\"unchen, 
Germany, and La Plata Observatory, Argentina}
\email{ahelmi@mpa-garching.mpg.de}

\author{Heather Morrison\altaffilmark{1,2,3}}
\affil{Astronomy Department, Case Western Reserve University, 
Cleveland, OH 44106}
\email{heather@vegemite.astr.cwru.edu}
\altaffiltext{2}{Cottrell Scholar of Research Corporation and NSF 
CAREER fellow}
\altaffiltext{3}{and Department of Physics}

\author{Mario Mateo\altaffilmark{1}}
\affil{Astronomy Department, University of Michigan, Ann Arbor, MI 
48109}
\email{mateo@astro.lsa.umich.edu}

\author{Edward W. Olszewski\altaffilmark{1}, Paul Harding\altaffilmark{1}}
\affil{Steward Observatory, University of Arizona, Tucson, AZ 85721}
\email{eolszewski@as.arizona.edu, harding@billabong.astr.cwru.edu}

\author{Kenneth C. Freeman, John Norris}
\affil{Research School of Astronomy \& Astrophysics, The Australian
National University, Mount Stromlo Observatory, Cotter Road, Weston
ACT 2611}
\email{kcf@mso.anu.edu.au, jen@mso.anu.edu.au}

\author{Stephen A. Shectman}
\affil{Carnegie Observatories, 813 Santa Barbara Street, Pasadena, CA 
91101}
\email{shec@ociw.edu}

\altaffiltext{1}{Visiting Astronomer, Kitt Peak National Observatory, 
National Optical Astronomy Observatories, which is operated by the 
Association of Universities for Research in Astronomy, Inc. (AURA) 
under the cooperative agreement with the National Science Foundation.}

\begin{abstract}

As part of the Spaghetti Project Survey (SPS) we have detected a
concentration of giant stars well above expectations for a smooth halo
model.  The position ($l\sim350, b\sim50$) and distance ($\sim 50$
kpc) of this concentration match those of the Northern over-density
detected by SDSS \citep{yan00, ive00}.  We find additional evidence
for structure at $\sim 80$ kpc in the same direction.  We present
radial velocities for many of these stars, including the first
published results from the 6.5m Magellan telescope.  The radial
velocities for stars in these structures are in excellent agreement
with models of the dynamical evolution of the Sgr dwarf tidal debris,
whose center is $60\degr$ away.  The metallicity of stars in these
streams is lower than that of the main body of the Sgr dwarf, which
may indicate a radial metallicity gradient prior to disruption.
    
\end{abstract}


\section{Introduction}

Decisive evidence is mounting in support of a Galactic halo 
predominately formed through mergers and accretion.  High-redshift 
observations, and modern simulations of structure evolution within the 
framework of cold dark matter \citep{pea99,ste99}, suggest that large 
galaxies formed hierarchically through the progressive merger of 
smaller pre-galactic structures.  Observational evidence of this 
process can be identified in fossil remains of past accretions or 
mergers in the Milky Way.

The remains of an accreted galaxy can be identified in the Galactic 
halo as coherent substructure in space \citep{kvj96}, velocity, 
or both \citep{hw99, har01}.  There are several examples of such 
coherent groups \citep{maj94, cot93, arn92, hmi99}.  One of the most 
dramatic is the Sgr dwarf \citep{iba94} and its extension to at least 
17 kpc to the SE \citep{mat98}, which demonstrates that galaxy 
accretion is a process that continues even today.

More recently \citet{yan00} and \citet{ive00} found an over-density of
blue horizontal branch (BHB) stars and RR Lyrae type stars,
respectively, in Sloan Digital Sky Survey (SDSS) commissioning data.
The over-density covers $35\degr$ on the sky and is located near
$\alpha = 14.7$ h, $\delta = 0\degr$ ($l=351\degr$, $b=52$), 50 kpc
from the Sun, and is $60\degr$ from the center of the Sgr dwarf.  This
structure has also been detected with carbon stars \citep{iba01}, and
CMD analysis \citep{mar01}. Very recently, this feature was also found
by \citet{viv01} as part of the QUEST RR Lyae type variable star
survey. They also found an over-density of RR Lyrae stars at closer
distances, partially associated with the globular cluster Pal 5, and
partially without identification.

Comparison to models of the disruption of the Sagittarius galaxy
\citep{kvj99, hw01, iba01} strongly suggests that this structure is
tidally stripped material from the Sgr dwarf \citep{ive00}. However,
none of the techniques to date have been able to confirm this with
both distance and velocity data.  As part of the Spaghetti Project
Survey (SPS) \citep{mor00}, we have serendipitously identified giants
associated with the SDSS over-density, and measured their radial
velocities and distances.  The excellent agreement with model
predictions leads us to conclude that this structure is, indeed, tidal
debris from the Sgr dwarf. Furthermore, we have identified additional
structures at different distances which may be multiple wraps of the
Sgr dwarf tidal stream.

\section{The Spaghetti Survey}

The SPS is a photometric and spectroscopic survey designed to identify 
structure in the Galactic halo \citep{mor00}.  For this study we use 
the modified Washington photometric system \citep{can76, gei84} to 
identify candidate red giants.  The $M-51$ color index is sensitive to 
surface gravity, while $C-M$ gives a photometric abundance.  Candidate 
halo stars MUST be observed spectroscopically to confirm the 
photometric classification and metallicity, and to obtain a radial 
velocity.

By chance, a number of survey fields with existing photometry and 
spectroscopy lie near the direction of the SDSS over-density.  We will 
focus on 16 of these fields which lie in the region defined by $300 < 
l < 360$, $0<l<30$, and $30<b<70$, and were imaged in April 1999 
\citep{doh00}.  A complete list of these fields, and the stellar 
photometry, is available through the NASA ADC database 
(http://adc.gsfc.nasa.gov).

We applied our giant selection criteria \citep{mor01} to the
photometry from this subset of fields.  In order to include some
fainter, potentially more distant giants in our first spectroscopic
observations with Magellan, we relaxed the $M-51$ error selection
limit, which is the most crucial \citep{mor01}, from 0.02 to 0.032
magnitudes, matching the largest error among all previously confirmed
giants in these fields.  There are 32 giant candidates, of which 21
have been confirmed spectroscopically using the criteria of
\citet{mor01}.  These are listed in Table \ref{tabgiants}.

The photometric metallicity \citep{mor00}, coupled with globular 
cluster giant branches \citet{dac90}, transformed to $M-T2$ 
\citep{mor01}, allows us to determine the absolute magnitude of each 
star.  The metallicity determined in this way is subject to errors of 
order 0.3 dex, which leads to distance errors of about 25\%.

The spectra for the six most distant giants are shown in Fig.\ 
\ref{figspectra}, along with two standards for comparison.  Star 
l355.89b+51.10 was observed on 20-22 February 2001 with the newly 
commissioned 6.5m Magellan I telescope at Las Campanas, with the LCO 
B\&C spectrograph.  Details of the spectral reduction process can be 
found in \citet{mat01}.  Both velocities \citep{mat01} and preliminary 
metallicities \citep{mor01} were determined from the spectra.  
Velocities are accurate to 20 km/s.  In most cases the agreement 
between the spectroscopic and the photometric metallicities is better 
than 0.2 dex.  The photometric value is listed in Table 
\ref{tabgiants}, with the exception of l356.15b+50.95, whose 
preliminary spectroscopic metallicity differed significantly from the 
photometric value.

\section{Model Comparison}

Fig.\ \ref{figlum} is a histogram of the heliocentric distance for all 
giant candidates.  The filled histogram is the subset of candidates 
that have been confirmed to be giants spectroscopically.  The curve is 
the predicted number of giants based on a model from \citet{mor93} for 
a smooth $R^{-3.5}$ halo. The model is normalized using the local halo 
giant density \citep{mor93}, and has an axial ratio variation 
prescribed by \citet{pre91}. A model was made for each of the 16 
selected fields, including a bright cutoff limit which varied from 
field to field, and a constant faint cutoff at $V=20$.

There is a concentration of candidates between 40 and 60 kpc, matching 
the distance of the SDSS over-density.  Of these, four stars have 
radial velocities.  Based on our success rate for stars at this 
magnitude, we expect approximately half the remaining candidates at 
this distance will be confirmed to be giants, and the other half will 
be revealed to be subdwarfs \citep{mor01}.  Taking this into 
account, there remain 3-4 stars in each of the three bins near 50 kpc, 
where we only expect 1-2 stars.

Even more striking is the correlation of radial velocities for stars 
in this structure.  The velocities of the four stars with spectra are 
remarkably similar ($\sigma = 31.5$ km/s) compared to the velocity 
dispersion of all confirmed giants ($\sigma_{\rm obs} = 150.2$ km/s).  
This association is certainly indicative of a coherent structure.

The models developed by \citet{hw01} predict that the Sgr dwarf
corresponds to only the central region of a much larger, at least a
few times $10^8$ M$_\odot$, progenitor.  These models predict that a
large amount of mass would be expected in streams, which can either be
stellar or dark-matter dominated.  In Fig.\ \ref{figmodel} we plot the
heliocentric distance and radial velocity for the particles in their
stellar model which fall within our selected region of the sky.  We
have also plotted the locations of all candidate giants with measured
radial velocities.  The four stars near 50 kpc are shown with filled
circles, and match very well both the distance and radial velocity
found in the models.

In addition to the concentration of giant candidates near 50 kpc,
there is one near 20 kpc and one near 80 kpc (Fig.\ \ref{figlum}).
The radial velocities of most of the stars near 20 kpc (Fig.\
\ref{figmodel}) match the predictions for the Sgr dwarf streams,
however, the range of predicted velocities at this distance is so
large that this cannot be considered strong support for these being
associated with the Sgr dwarf tidal debris.  It is interesting to note
that if most of these 20 kpc stars are indeed Sgr debris, then the
smooth halo density at this distance must be much lower than simple
models predict.  This would suggest a large fraction of the halo may
be composed of stream-like structures, even as close as $R\sim 20$
kpc.

In contrast to the $\sim 20$ kpc concentration, the two stars at 80
kpc show a spatial density and a velocity correlation above that
expected from a smooth halo.  Their position and velocity match model
predictions for an earlier ``wrap'' of the Sgr dwarf tidal stream.

We compared the distance and velocity distribution of the 21
candidates with that of a smooth halo model. The smooth halo has an
$R^{-3.5}$ density profile, and a radially anisotropic velocity
ellipsoid with $\sigma = 135 $ km/s.  We performed 10,000 Monte Carlo
simulations of 21 stars drawn from the smooth halo, including
observational errors. The fraction of simulations which gave the
observed distribution of 4 stars near 50 kpc with velocity dispersion
of 32 km/s, {\em and} 2 stars near 80 kpc with velocity dispersion 21
km/s, was 11 in 10,000. We also performed this same exercise by
drawing the samples from the model of the Sgr dwarf (Helmi \& White
2001). Nearly 70\% of these 10,000 samples result in the observed
distribution.  Thus, the likelihood that these stars belong to a
smooth halo is negligible.

The present data cannot be used to make a full comparison to the
spatial density of the model stream. In particular, the spatial
sampling of our fields, at this point, is too sparse to determine the
width or direction of the stream, and numerous selection effects must
be addressed. We can only note that the confirmed members at 50 kpc
are spread over $16\degr$ in longitude. If the two stars at 80 kpc are
included the spread is over $50\degr$ in longitude. Furthermore, a
detailed density comparison, for example, to determine if the 20 kpc
stars are part of a stream or part of a smooth halo, must await a
larger area to be studied, preferentially located far away from the
expected sky position of the Sgr streams \citep{hel01}.

Finally, we note that the mean metallicity of the six stars we claim
to be part of the Sgr stream ([Fe/H] $\sim -1.5$) is about 0.5 dex
lower than the mean for field stars in the main body of the Sgr galaxy
\citep{lay97, mat98}.  Many dSph galaxies show radial gradients in
their horizontal branch (HB) morphologies, such that the outer HB
stars are bluer \citep{cal98, hur99, dac00, harb01}.  Since the
outermost stars are preferentially stripped during tidal disruption
\citep{pia95, oh95}, the streams could exhibit a different HB
population than the more tightly-bound core.  This could explain why
the SDSS over-density consists of large numbers of blue HB stars,
while such stars are mostly absent in the core of Sgr. This is also
consistent with the lower metallicity of the stream stars if the HB
morphology gradient reflects an underlying metallicity variation in
Sgr.

The destruction of dwarf galaxies is probably a crucial element in
building up the stellar halo of our Galaxy.  With further
observations, the age information that we can recover from the
different wrappings, combined with the metal abundances of the stream
stars, will give a detailed observational picture of the progressive
destruction of this galaxy. Deriving the chemical enrichment and star
formation as a function of time and position in the Galaxy will, for
example, help us understand the effect of tides on the internal
evolution of such apparently fragile systems. Mapping the streams of
Sgr will also provide some strong constraints on the large scale
evolution of the shape and structure of the Galactic potential on Gyr
time-scales, and on the amount of dark matter substructure.

\acknowledgements

This work was supported by NSF grants AST 96-19490, AST 0098435 to
HLM, AST 95-28367, AST 96-19632, AST 98-20608, AST 0098661 to MM, AST
96-19524, AST 0098435 to EWO which partially supported EWO, RCDP, and
PH, and CONICET and Fundaci\'on Antorchas grants to AH. We wish to
thank the support teams at KPNO, CTIO, and Las Campanas for their help
in acquiring the data.  We are very grateful to the NOAO TAC for
their consistent strong support of this project through generous
telescope allocations.  We wish to offer a special thank you to the
Magellan Project for producing such a fine telescope and to Ian
Thompson for his help in obtaining SPS data during telescope
commissioning. Finally, we are grateful to the referee for insightful
comments that improved this work.

\begin{deluxetable}{ccccccc}
\tabletypesize{\scriptsize}
\tablecaption{Confirmed and Potential Giants Associated with Sgr Tidal 
Debris}
\tablewidth{0pt}
\tablehead{
\colhead{Star} & 
\colhead{$V_0$} & 
\colhead{$(M-T2)_0$} & 
\colhead{[Fe/H]\tablenotemark{a}} & 
\colhead{$M_V$} & 
\colhead{Dist. (kpc)} & 
\colhead{Helio. Vel. (km/s)}
}
\startdata
l003.06b+61.30 & 19.21 & 1.174 & -1.8 & 0.830 & 47.5 &   12 \\    
l011.85b+51.95 & 17.57 & 1.192 & -1.3 & 0.720 & 23.4 & -143 \\    
l017.04b+46.40 & 16.81 & 1.130 & -1.7 & 1.097 & 13.9 &   56 \\    
l017.35b+46.50 & 16.53 & 1.237 & -1.5 & 0.200 & 18.3 &  112 \\    
l301.78b+45.46 & 16.68 & 1.130 & -1.8 & 1.059 & 13.3 &      \\    
l302.36b+49.04 & 17.38 & 1.255 & -3.6 &-0.007 & 30.0 &      \\    
l302.43b+48.83 & 19.86 & 1.136 & -1.5 & 1.141 & 55.5 &      \\    
l304.49b+60.51 & 18.90 & 1.363 & -1.8 &-0.922 & 80.0 &   63 \\    
l304.69b+60.52 & 16.86 & 1.197 & -1.8 & 0.276 & 20.7 &  278 \\    
l305.22b+61.24 & 17.24 & 1.187 & -1.8 & 0.307 & 24.4 &   56 \\    
l305.32b+60.58 & 16.91 & 1.177 & -1.7 & 0.484 & 19.2 &  -62 \\    
l305.44b+61.34 & 16.38 & 1.178 & -1.1 & 0.938 & 12.3 &      \\    
l305.50b+60.65 & 17.52 & 1.370 & -1.0 &-0.204 & 35.1 &  207 \\    
l322.12b+39.91 & 16.18 & 1.180 & -1.4 & 0.772 & 12.1 &  329 \\    
l322.18b+40.02 & 19.51 & 1.120 & -2.2 & 0.889 & 53.1 &      \\    
l326.26b+49.00 & 17.79 & 1.236 & -1.6 & 0.176 & 33.4 &  -68 \\    
l332.71b+46.84 & 17.64 & 1.144 & -1.5 & 1.050 & 20.8 &  142 \\    
l333.34b+46.51 & 19.10 & 1.111 & -1.6 & 1.331 & 35.7 &      \\    
l333.50b+46.75 & 18.27 & 1.138 & -1.3 & 1.179 & 26.1 &   76 \\    
l338.85b+68.27 & 16.66 & 1.218 & -2.4 &-0.037 & 21.9 &  -28 \\    
l340.15b+68.30 & 19.49 & 1.169 & -1.5 & 0.823 & 54.1 &      \\    
l347.28b+53.30 & 19.06 & 1.163 & -1.7 & 0.882 & 43.2 &   84 \\    
l347.42b+53.31 & 16.50 & 1.208 & -1.5 & 0.437 & 16.3 &  169 \\    
l347.68b+53.06 & 17.02 & 1.121 & -1.8 & 1.107 & 15.2 & -119 \\    
l354.95b+66.01 & 19.86 & 1.149 & -1.6 & 0.992 & 59.3 &      \\    
l355.89b+51.10 & 19.86 & 1.232 & -1.4 & 0.346 & 80.0 &   33 \\     
l355.99b+51.16 & 16.90 & 1.187 & -1.5 & 0.626 & 18.0 & -113 \\    
l356.15b+50.95 & 18.26 & 1.304 & -1.5 &-0.290 & 51.1 &   47 \\    
l356.54b+51.18 & 19.10 & 1.221 & -1.0 & 0.724 & 47.3 &      \\    
l356.70b+51.23 & 17.16 & 1.446 & -1.6 &-1.072 & 44.2 &   25 \\    
l356.81b+51.06 & 19.28 & 1.146 & -1.4 & 1.088 & 43.5 &      \\    
l356.88b+51.09 & 19.14 & 1.138 & -1.3 & 1.200 & 38.8 &      \\    
\enddata
\label{tabgiants}
\tablenotetext{a}{All metallicity determinations are photometric, except 
that for star l356.15b+50.95, which is a preliminary spectroscopic 
measurement.}
\end{deluxetable}

\begin{figure}
%\epsscale{0.7} 
%\plotone{Dohm-Palmer.fig1.ps} 
\caption{The spectra of the 6 stars beyond 40 kpc, and two standard 
stars.  G194-37 is a known subdwarf with [Fe/H]=-2.0, and N4590-71 
is a globular cluster giant with [Fe/H]=-2.1.  The tick marks 
indicate the zero flux level for each successive star.  The dotted 
lines mark three spectral indicators used to distinguish dwarfs from 
giants: Ca II K (3934\AA), Ca I (4227\AA), and Mg b (5167\AA).  Note 
that the subdwarf has a much stronger CaI line than any of the distant 
giants.}
\label{figspectra}
\end{figure}

\begin{figure}
%\epsscale{0.7} 
%\plotone{Dohm-Palmer.fig2.ps} 
\caption{A histogram of giant candidates in heliocentric distance.  
The giants stars come from 16 selected fields with Galactic 
coordinates $300<l<360, 0<l<30, and 30<b<70$.  The shaded histogram 
shows all spectroscopically confirmed giants.  To match the largest 
error of a confirmed giant, we have included all candidates with a 
$M-51$ error $<0.032$.  Given this error limit, we expect 
approximately half the unconfirmed stars near 50 kpc are actually 
metal-poor subdwarfs.  The solid line is a model prediction from a 
smooth $R^{-3.5}$ density profile, based on the selected fields 
observed.  Note the over-densities at 50 and 80 kpc, and 
possibly near 20 kpc.}
\label{figlum}
\end{figure}

\begin{figure}
%\epsscale{0.7} 
%\plotone{Dohm-Palmer.fig3.ps} 
\caption{Distance versus radial velocity for the Sgr dwarf stellar
models of \citet{hw01}.  The model points come from the range
$300<l<360, 0<l<30$. In the top panels they have been divided into
three latitude bins, while all latitudes are included in the bottom
panels.  Also plotted in the top panels are the giants with measured
radial velocity.  The diamonds mark the two most distant giants at 80
kpc, the filled circles mark the 4 giants near 50 kpc, the triangles
mark stars matching the model near 20 kpc, and the open circles mark
stars that don't match the model within their error box. The
bottom left plot shows the same data as the top panel, except not split
into latitude bins. For comparison, we plot in the bottom right one of
the 10,000 Monte Carlo simulations of 21 stars drawn from a smooth
halo population.}
\label{figmodel}
\end{figure}

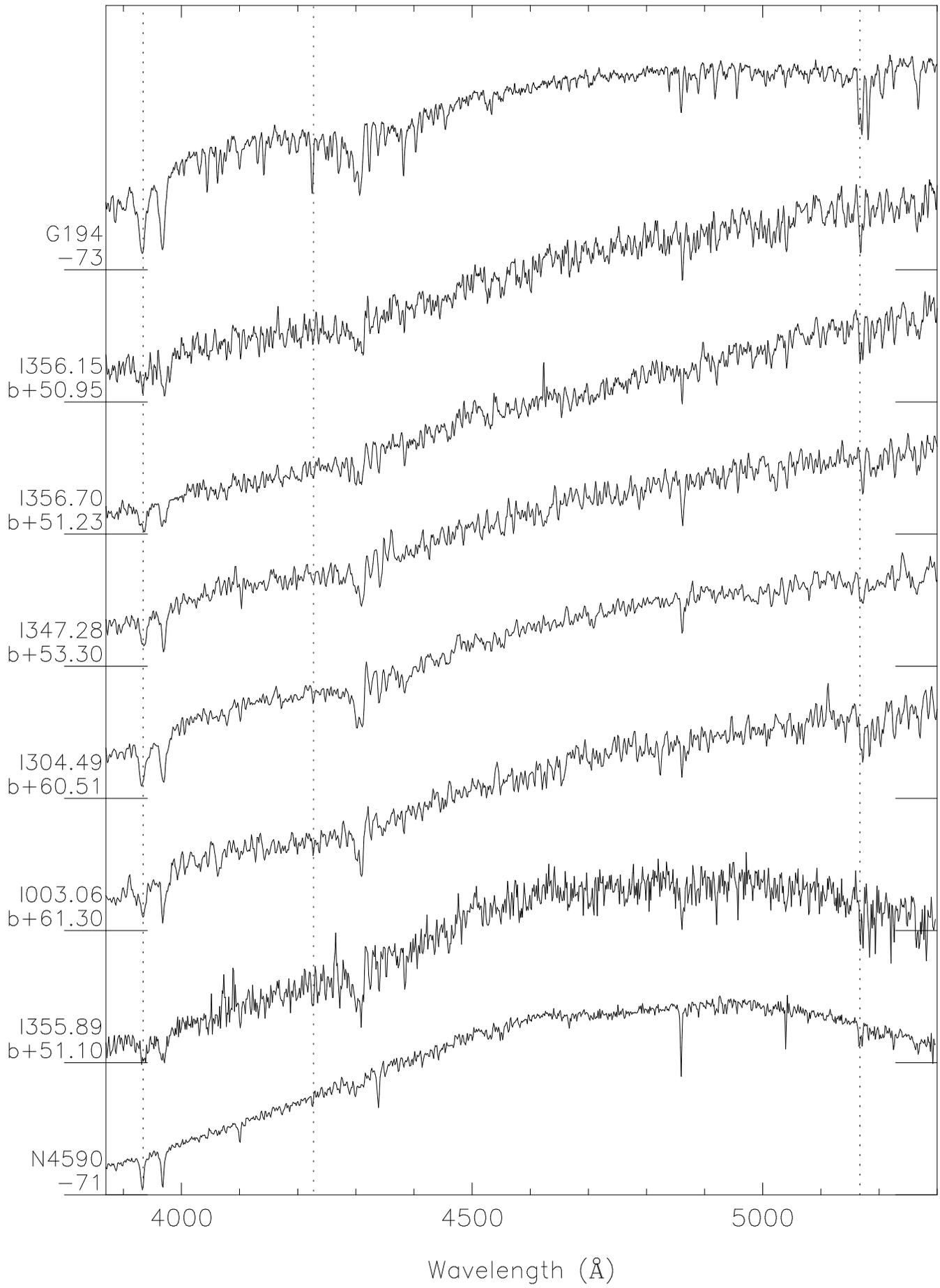
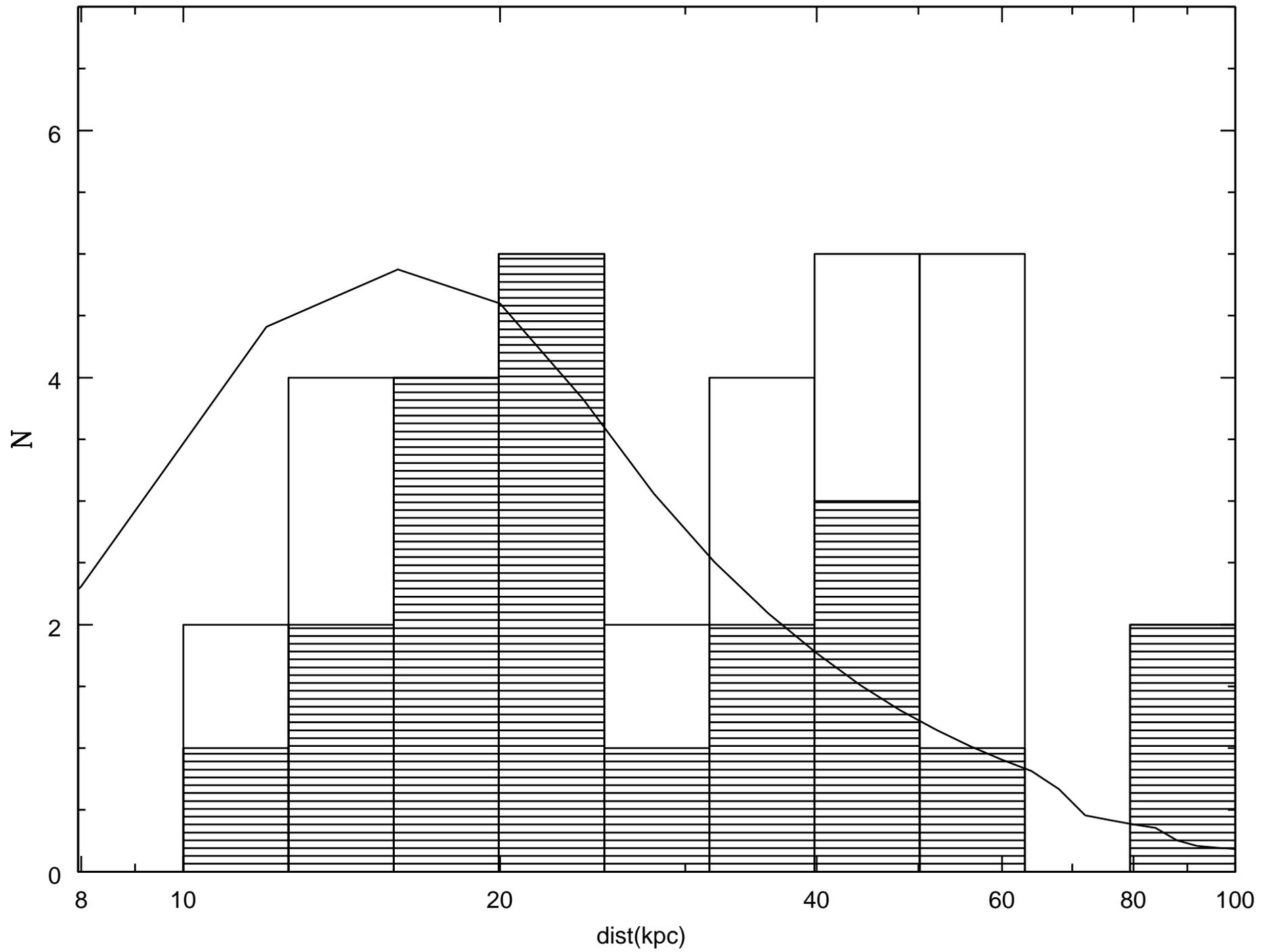
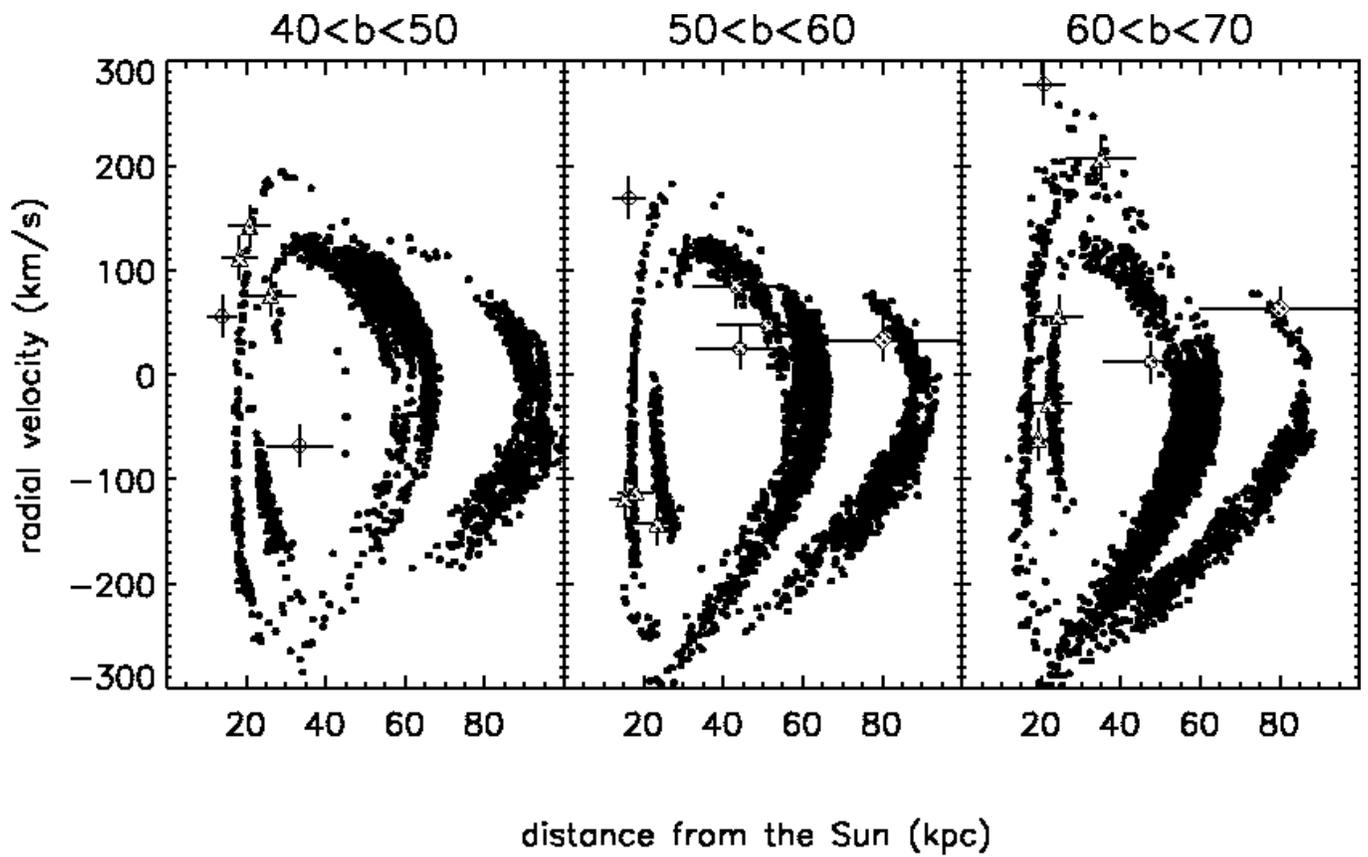
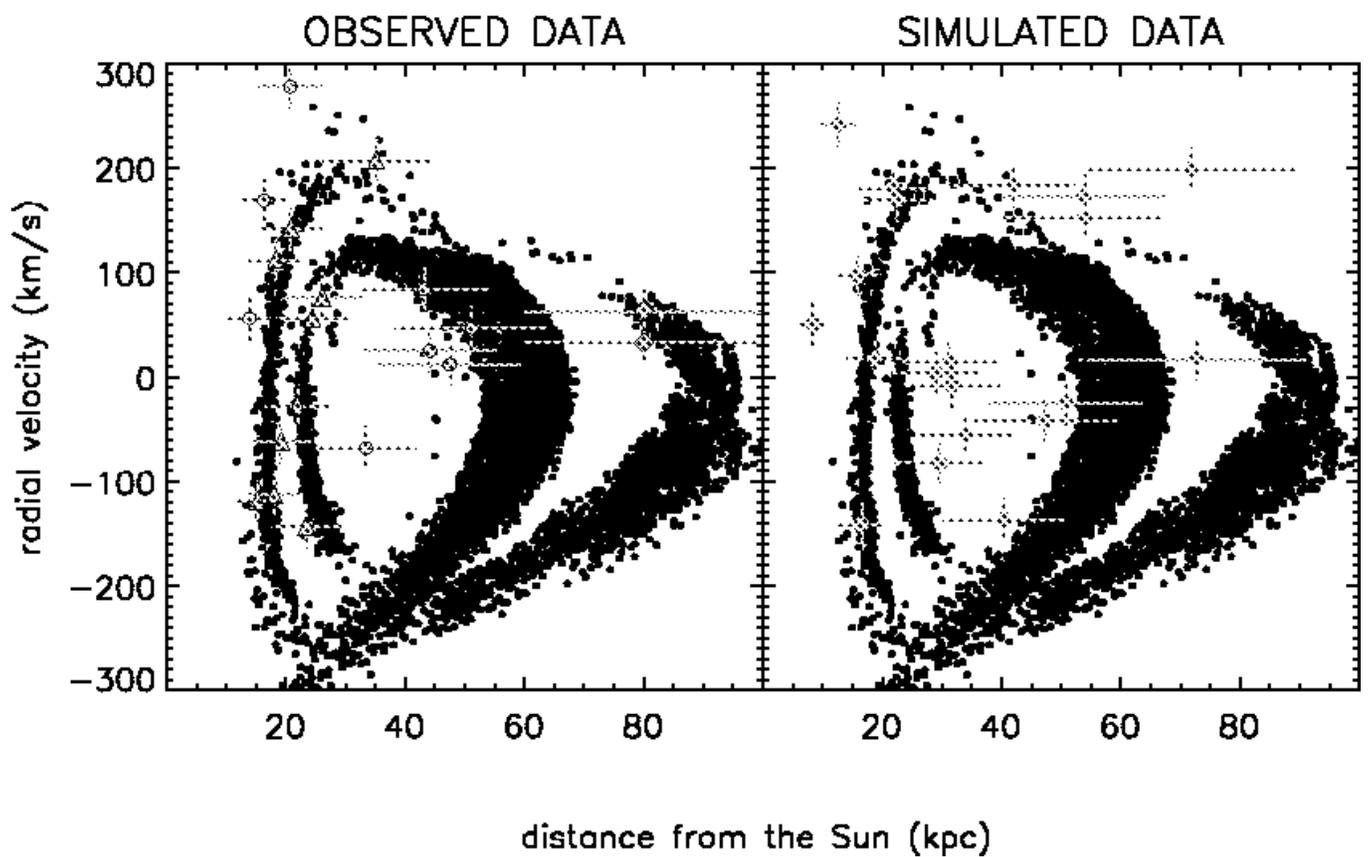
\end{document}